\documentclass{PoS}
\usepackage{amsfonts,amsmath,amssymb}
\usepackage{mathtools}
\usepackage{graphics}

\newcommand{\csw}{c_\mathrm{SW}}

\usepackage{multicol} 
\newcommand{\be}{\begin{equation}}
\newcommand{\ee}{\end{equation}}
\newcommand{\bea}{\begin{eqnarray}}
\newcommand{\eea}{\end{eqnarray}}
\newcommand{\nn}{\nonumber}

\title{ Quark masses and decay constants in $N_f=2+1+1$ isoQCD with Wilson clover twisted mass fermions}

\ShortTitle{}

\author{G.~Bergner$^a$, P.~Dimopoulos$^b$, J.~Finkenrath$^c$, E.~Fiorenza$^d$, R.~Frezzotti$^e$, \,\, \speaker{M.~Garofalo}~\thanks{marco.garofalo@roma2.infn.it}\,~$^f$,
B.~Kostrzewa$^g$,
 F.~Sanfilippo$^{h}$, S.~Simula$^h$,
U.~Wenger$^i$  \\
\vspace{-0.5cm}
\begin{center}\mbox{\includegraphics[scale=0.02]{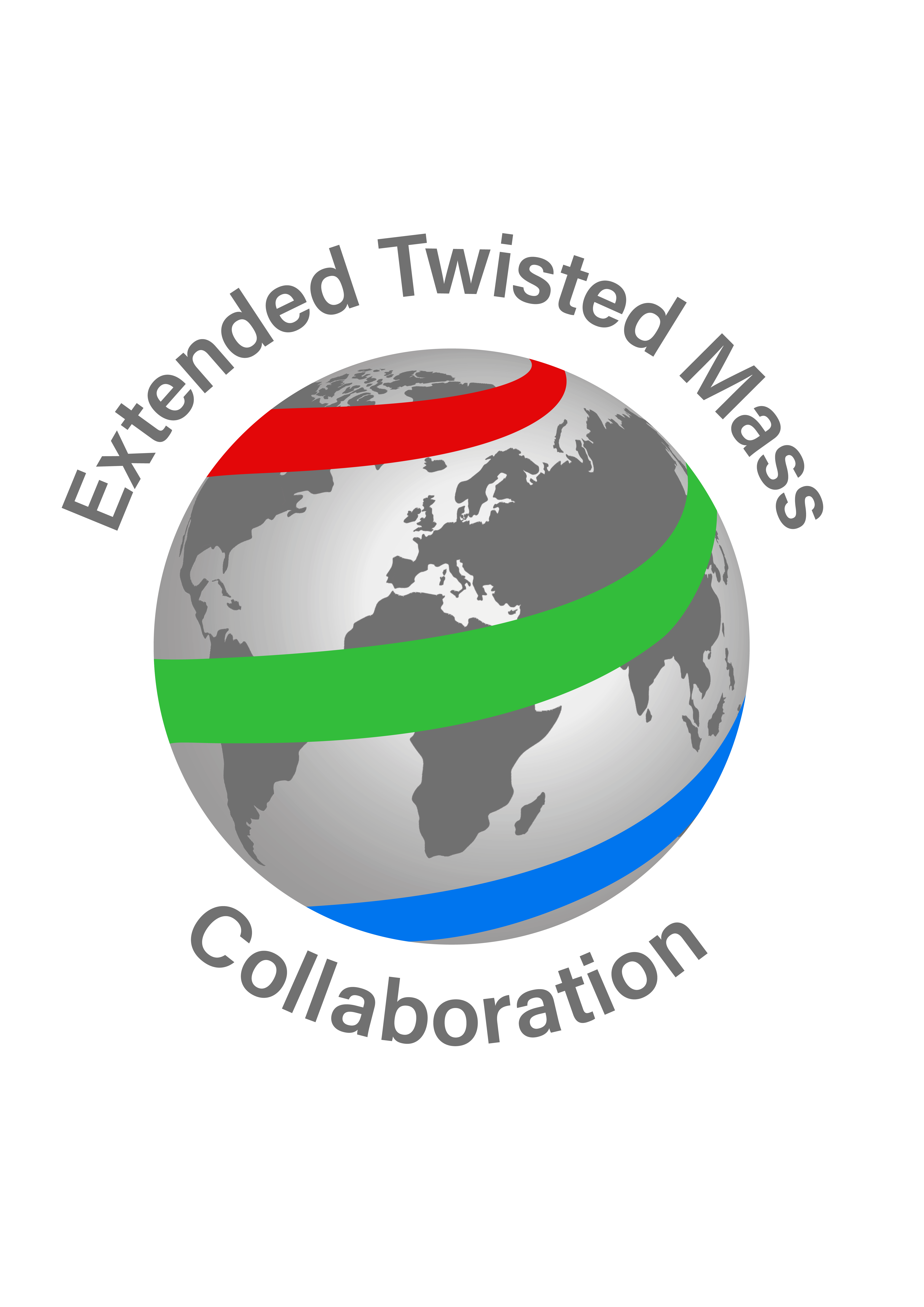}}
\end{center}\\
\llap{$^a$} University of Jena, Institute for Theoretical Physics,Max-Wien-Platz 1, D-07743 Jena, Germany\\
\llap{$^b$}Dipartimento di Scienze Matematiche, Fisiche e Informatiche, Università di Parma, via Università, 12 - I 43121 Parma\\
\llap{$^c$}Computation-based Science and Technology Research Center,
The Cyprus Institute, 20 Konstantinou Kavafi Street, 2121 Nicosia, Cyprus\\
\llap{$^d$} CP3 -Origins \& IMADA, University of Southern Denmark, Campusvej 55, 5230 Odense, Denmark\\
\llap{$^e$} University of Rome Tor Vergata and INFN Roma Tor Vergata, Via della Ricerca Scientifica 1,I-00133, Rome, Italy\\
\llap{$^f$}Dip. di Matematica e Fisica, University of Roma Tre,Via della Vasca Navale 84, I-00146 Rome, Italy\\
\llap{$^g$}HISKP (Theory), Rheinische Friedrich-Wilhelms-Universit\"at Bonn, Nussallee 14-16, 53115 Bonn, Germany\\
\llap{$^h$} INFN section of Roma Tre, Via della Vasca Navale 84, I-00146 Rome, Italy \\
\llap{$^i$} Institute for Theoretical Physics, Albert Einstein Center for Fundamental Physics, University of Bern, Sidlerstr. 5, CH-3012 Bern, Switzerland    
}
\abstract{We present a preliminary study of the pion, kaon and D-meson masses and decay constants
in isosymmetric QCD, as well as a preliminary result for the light-quark renormalized mass. The analysis is based on the gauge ensembles produced by ETMC with $N_f=2+1+1$ flavours of Wilson-clover twisted mass quarks, spanning a range of lattice spacings from $\sim0.10$ to $0.07$ fm and include configurations at the physical pion point on lattices with linear size up to $L~\sim~5.6$~fm}

\FullConference{37th International Symposium on Lattice Field Theory - Lattice2019\\
		16-22 June 2019\\
		Wuhan, China}

\begin{document}
 \vspace*{-1.cm}
\section{Action}\vspace{-0.3cm}
\label{sec:action}

We present a preliminary analysis based on the gauge ensembles produced by the Extended Twisted Mass Collaboration (ETMC) in isosymmetric QCD (isoQCD) with $N_f=2+1+1$ flavours of Wilson-clover twisted mass quarks \cite{Alexandrou:2018egz}, i.e.\ in a lattice setup where physical observables can be evaluated at maximal twist with no O($a$) scaling violations~\cite{Frezzotti:2003ni}. 
The lattice action is given by 
\begin{equation}
    S = S_g + S_{tm}^\ell  + S_{tm}^h ~ ,
    \label{eq:totalaction}
\end{equation}
where for $S_g$  we choose the Iwasaki improved gluon  action (see e.g. eq.~(2) of \cite{Alexandrou:2018egz}).
For the light (up and down) sea quark doublet $\chi_\ell=(u,d)^t$, 
the twisted mass action~\cite{Frezzotti:2000nk} takes the form
\begin{equation}
  \label{eq:sflight}
  S_{tm}^\ell = \sum_x \bar{\chi}_\ell(x)\left[ D_W(U) + \frac{i}{4} \csw \sigma_{\mu\nu}
    \mathcal{F}_{\mu\nu}(U) + m_0 + i \mu_\ell \tau^3 \gamma^5 
  \right] \chi_\ell(x)\,.
\end{equation}
where $\mu_\ell$ is the twisted and 
$m_0$ the (untwisted) Wilson bare quark mass. The Pauli matrix $\tau_3$ acts in flavour 
space and $D_W$ is the massless Wilson-Dirac operator. The Wilson quark mass $m_0$ 
and the clover term $\frac{i}{4} \csw \sigma_{\mu\nu}\mathcal{F}_{\mu\nu}(U)$ with 
Sheikoleslami-Wohlert improvement coefficient  $c_\mathrm{SW}$ \cite{Sheikholeslami:1985ij}
are trivial in flavour space. 

For the strange ($s$) and charm ($c$) sea quark pair (field $\chi_h=(c,s)^t$) 
the action reads~\cite{Frezzotti:2003xj}
\begin{equation}
  \label{eq:sfheavy}
  S_{tm}^h = \sum_x \bar{\chi}_h(x)\left[ D_W(U) + \frac{i}{4} \csw \sigma_{\mu\nu}
    \mathcal{F}_{\mu\nu}(U) + m_{0h} - \mu_\delta \tau_1 + i \mu_\sigma \tau^3 \gamma^5
  \right] \chi_h(x)\,.
\end{equation}
The term $\mu_\delta \tau_1$ is absent in eq.~(\ref{eq:sflight}) as the $u$ and $d$ quarks 
are mass degenerate. 
By tuning the light and heavy Wilson bare quark masses  $m_0$ and $m_{0h}$  to their common critical value $m_{crit}$  the maximally twisted fermion action is obtained for which all 
physical observables are $\rm{O}(a)$-improved \cite{Frezzotti:2003ni,Frezzotti:2003xj}.

In this framework isospin breaking lattice artifacts affect
significantly the unitary neutral
pion mass making it typically smaller than its charged counterpart,
which in turn may
render unquenched Monte Carlo simulations numerically unstable. This
phenomenon here is
substantially suppressed by introducing a clover term in the action~\cite{Becirevic:2006ii,Alexandrou:2018egz}, which
proves crucial for simulations close to the physical pion 
point with lattice
spacings in the range (0.10, 0.07)~fm~\cite{Alexandrou:2018egz}, where topological freezing is
not  a problem.
Within a $\sim 10\%$ accuracy, the clover term coefficient $c_{SW}$ can be fixed through 
its estimate in one-loop \cite{AOKI1999501} tadpole boosted perturbation theory, namely
$
c_{SW} \cong 1+ 0.113(3) \frac{g_0^2}{P} 
$
with $P$ the plaquette expectation value. Here we follow this prescription at 
all values of $g_0^2$ (corresponding to lattice spacings $a \sim 0.10,\, 0.08,\, 0.07$ 
fm - see Sect.~2). 
The parameters of the various simulation ensembles are shown in Table~\ref{ensembles}.

For the $(c,s)$ quark sector we adopt a mixed action setup using Osterwalder-Seiler 
fermions in the valence, with the same critical mass, $m_{crit}$, as determined in the 
unitary setup and with action~\cite{Frezzotti:2004wz}
\begin{equation}
  S_{tmOS}^{h,val} = \sum_{f=c,s}\sum_x \bar{\chi}_f(x)\left[ D_W(U) + \frac{i}{4} \csw \sigma_{\mu\nu}
    \mathcal{F}_{\mu\nu}(U) + m_{crit}  + i \mu_f^{OS}  \gamma^5
  \right] \chi_f(x)\,.
\end{equation}
Reflection positivity of renormalized correlation functions in the continuum limit 
is guaranteed because the renormalized $c$ and $s$ valence masses are matched to their sea
counterparts through
\begin{equation}
m^{val,ren}_{c,s}=\frac{1}{Z_P}\left(\mu_\sigma\pm \frac{Z_P}{Z_S}\mu_\delta  \right) \; ,
\end{equation} 
with $Z_P$ and $Z_S$ denoting the non-singlet pseudoscalar and scalar Wilson fermion quark 
bilinear renormalization constants. In this way we avoid any undesired $s$-$c$ quark 
mixing (through cutoff effects) in the valence and preserve the automatic O($a$) improvement
of physical observables~\cite{Frezzotti:2004wz}.

The masses of the sea strange and charm quarks are set, at each $\beta$, to values 
close to the physical ones (with a few percent tolerance) through the tuning procedure
described in Ref.~\cite{Alexandrou:2018egz}. As for valence mass parameters, we evaluated 
correlators at light valence quark mass $\mu_\ell$ equal to its sea counterpart, as well as 
at three values of quark mass ($\mu_s$) in the strange region and four values of the quark 
mass ($\mu_c$) in the charm region (see Table~\ref{valence_masses}), which allows for a 
precise interpolation
to the physical $s$ and $c$ point as determined by the $K$- and $D$-meson masses.
Quark propagators with light and strange-like valence masses are obtained from inversions of the Dirac Matrix using the DD$\alpha$AMG multi-grid algorithm optimized for 
twisted mass fermions~\cite{Alexandrou:2016izb}.

\begin{table}\centering\small
\begin{tabular}{|c|ccccccc|}
\hline
 name &$L^3\cdot T/a^4$& Nconf  &$\kappa$ &$a\mu_\ell$ &$a\mu_\sigma$& $a\mu_\delta$ &  
$\beta$ \\
 \hline
cA211.53.24 &$24\cdot 48$& 628 &0.1400645 &0.0053& 0.1408 & 0.1521&1.726   \\
             \hline
cA211.40.24 &$24\cdot 48$& 662 &0.1400645 &0.0040 &0.1408 & 0.1521&1.726 \\
 \hline
cA211.30.32 &$32\cdot 64$& 1237 &0.1400645   & 0.0030&0.1408 & 0.1521 &1.726\\
\hline
cA211.12.48 &$48\cdot 96$&  322 &0.1400650 &0.0012 &0.1408 & 0.1521   &1.726 \\\hline
cB211.25.48 &$48\cdot 96$& 314  & 0.1394267 & 0.0025 &0.1246864 &0.1315052 &1.778 \\
 \hline
cB211.072.64&$64\cdot 128$ & 187 & 0.1394265 & 0.00072 &0.1246864 &0.1315052 &1.778 \\
  \hline
cC211.06.80 &$80\cdot 160$&  210  &0.1387510 & 0.0006 &0.106586 & 0.107146 & 1.836 \\   
\hline
\end{tabular}
\caption{Simulations details for the ETMC gauge ensembles with $N_f=2+1+1$ Wilson-clover 
twisted mass quarks: volume, number of gauge
configurations analyzed, $\kappa = (8 + 2 am_{crit})^{-1}$,  
bare twisted masses $a\mu_\ell$ , $a\mu_\sigma$, $a\mu_\delta$,
and $\beta$. We have set $c_{SW}=[1.74,1.69,1.645]$ for $\beta=[1.726,1.778,1.836]$, 
respectively. }\label{ensembles}
\end{table}

\begin{table}\centering\small
\begin{tabular}{|c|ccc|}
\hline
 name & $a\mu_\ell^{valence}=a\mu_{\ell}$ &$a\mu_s$& 	$a\mu_c$  \\
 \hline
cA211.53.24 & 0.0053  &0.0176,  0.0220, 0.0264&      0.2596, 0.2856, 0.3115, 0.3433 \\
             \hline
cA211.40.24 & 0.0040  &0.0176,  0.0220, 0.0264&      0.2596, 0.2856, 0.3115, 0.3433\\
 \hline
cA211.30.32 & 0.0030  &0.0176,  0.0220, 0.0264&      0.2596, 0.2856, 0.3115, 0.3433\\
\hline
cA211.12.48 & 0.0012  &0.0176,  0.0220, 0.0264&      0.2596, 0.2856, 0.3115, 0.3433\\
\hline
cB211.25.48 & 0.0025 & 0.0148,  0.0185,  0.0222&     0.2181,  0.2399,  0.2617,  0.2884\\
 \hline
cB211.072.64& 0.00072 & 0.0148,  0.0185,  0.0222&    0.2181,  0.2399,  0.2617,  0.2884\\
  \hline
cC211.06.80 &0.0006  &  0.0128,  0.0161,  0.0193&    0.1907,  0.2098,  0.2288,  0.2522 \\
\hline
\end{tabular}
\caption{Values of the valence bare quark masses for each of the ensembles analysed here.
}\label{valence_masses}
\end{table}

For each ensemble we computed the two-point pseudo-scalar (PS) correlators defined as
 \be
    C(t) = \frac{1}{L^3} \sum\limits_{\vec{x}, \vec{z}} \left\langle 0 \right|P_{qq'} (x) P_{qq'} (z)^\dag \left| 0 \right\rangle \delta_{t, (t_x  - t_z )} ~ ,
    \label{eq:P5}
 \ee
where $P_{qq'} (x) = \overline{\chi}_q(x) i\gamma_5  \chi_{q'}(x)$ with 
single-flavour $\chi_q$ and $q$ in  $\{l,s,c\}$.
In this work for all mesons the twisted mass of the two valence quarks $q, \; q'$ are always taken with opposite signs (or equivalently in the physical quark basis of Ref.~\cite{Frezzotti:2003ni} the Wilson parameters of the two valence quarks take opposite values), as this choice is known to suppress O($a^2$) 
errors. These two-point correlators are evaluated for the four combinations resulting from smeared or local interpolating fields at the sink and/or the source and analysed through the GEVP method~\cite{Blossier:2009kd} for the extraction of the ground-state masses ($M_{PS}$) and matrix elements.~\footnote{As discussed in \cite{Blossier:2009kd} the mass of the lightest state is estimated through a $t$-plateau average of the smallest eigenvalue obtained (for a suitable choice of $t_0$) from the GEVP method, viz.\ 
$ \lambda_0(t,t_0)=C \left(e^{ - M_{PS}  (t-t_0)}  + e^{ - M_{PS}  (T - (t-t_0))}\right)$.} 
We employ a Jacobi smearing of the quark fields \cite{PhysRevD.47.5128} combined with  APE smearing of the gauge links \cite{FALCIONI1985624}.

For maximally twisted quarks the value of the matrix elements 
$\mathcal{Z}_{PS} = | \langle PS | P_{qq'}  | 0 \rangle|^2$ 
determines the PS-meson decay constant with no need of any renormalization 
constant~\cite{Frezzotti:2000nk}, namely
 \be
    af_{PS} = a (\mu_q+\mu_{q'}) \frac{\sqrt{a^4 \mathcal{Z}_{PS}}}{aM_{PS} \mbox{sinh}(aM_{PS})} ~ .
    \label{eq:decaypi}
 \ee

For all ensembles the value of the bare parameter $am_{0(h)}$ is tuned towards $am_{crit}$ so
that the renormalized untwisted current quark mass $Z_A am_{PCAC}$ is well below  
$0.1 a\mu_\ell$, which is enough to make negligible the O($a$) errors
due to small numerical deviations from maximal twist. A marginal exception occurs only
for the ensemble cA211.12.48, where a longer autocorrelation of $am_{PCAC}$ is observed 
in the MC simulation and $Z_A am_{PCAC} = -0.00015(4)$ for $a\mu_\ell=0.0012$ is found. 
This small systematic error has been corrected by ``reweighting''  
from $\kappa=0.1400650$ to $\kappa_{crit.}=0.1400640$. 
 \vspace{-0.5cm}
\section{Lattice calibration and determination of $w_0$ from $f_\pi$}
\begin{figure}[h]\centering
   \vspace{-0.6cm}
 \includegraphics[scale=0.5]{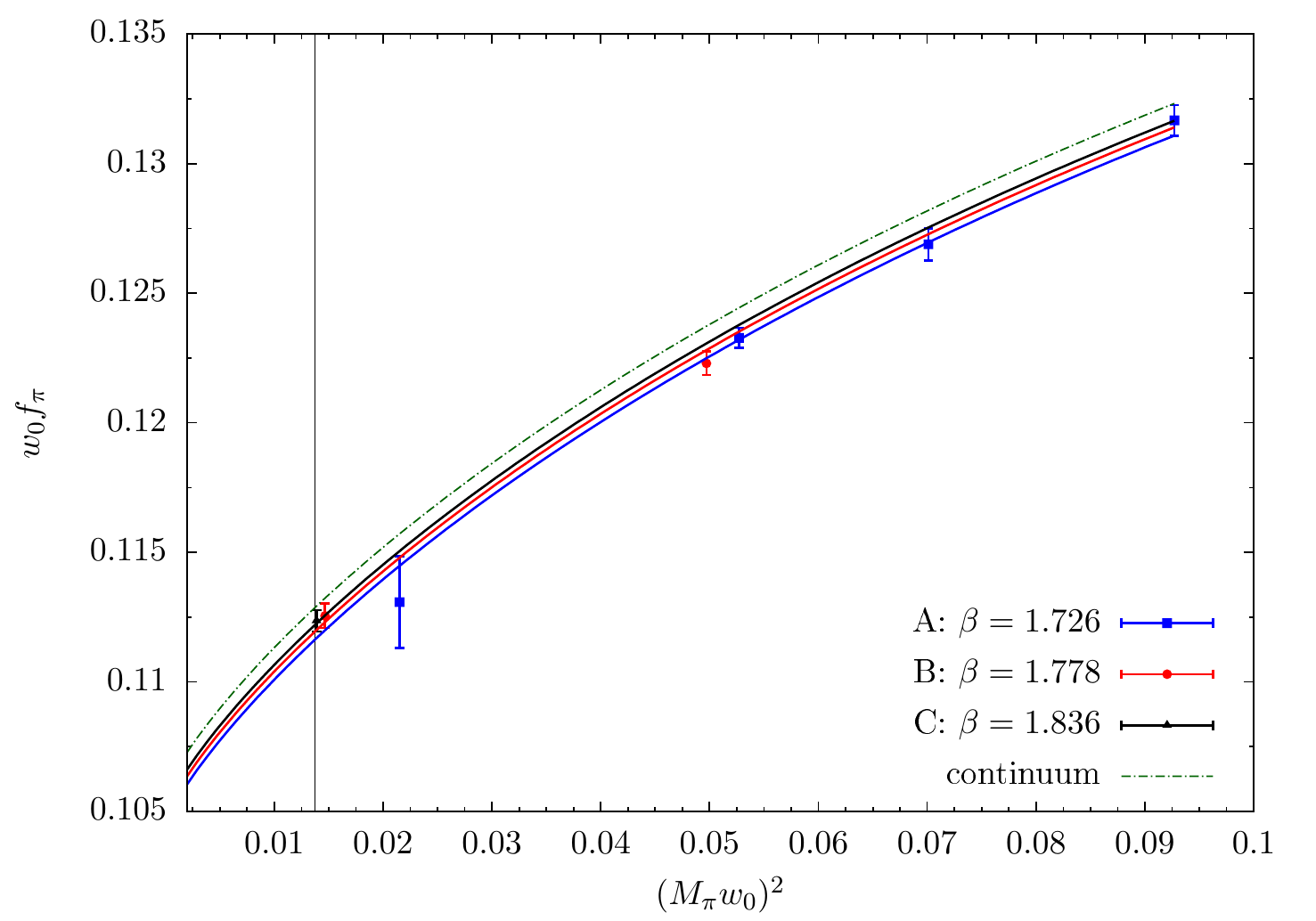}
  \vspace{-0.4cm}
 \caption{Preliminary chiral and continuum fit of $f_\pi$ as a function of $M_\pi$ in units of the gradient flow scale  $w_0$ \cite{Borsanyi2012}. The fit function we employ is given in Eq.~(\ref{f_pi_of_M_pi}) }
    \vspace{-0.3cm}
 \label{fig:f_pi_of_M_pi}
 \end{figure}

\vspace{-0.3cm}
We carried out fits of the dependence of $f_\pi$, written in units of the gradient flow scale 
$w_0$ \cite{Borsanyi2012}, on the meson mass $M_\pi^2$ using the  SU(2) chiral perturbation 
theory (ChPT) formula
 \bea\label{f_pi_of_M_pi}
\hspace{-0.8cm}    (f_\pi  w_0 ) & = &\!\!\!\! (f w_0 )\left[ 1 - 2\xi_\ell^M \log \xi_\ell^M  + P_3\xi_\ell^M  +  P_4  \,a^2/w_0^2 \right] K_f^{FSE} ~ ,
 \eea
 where $\xi_\ell^M=M_\pi^2/(16\pi^2f^2)$ and $f$, which has been left free to vary in the fit, is the SU(2) low-energy-constant entering the leading order chiral effective Lagrangian. The parameter $P_3$ is related to the next-to-leading low-energy-constant 
 $\bar{\ell_4}$ with
$ 
P_3=2\bar{\ell_4}+4\log\left( M_\pi^{phys}/(4\pi f) \right)\,,
 $
 with $M_\pi^{phys}$  being the value of the pion mass at the physical point.
The factor $K_f^{FSE}$ represents the correction for finite
size effects (FSE), as computed in Ref~\cite{Colangelo:2005gd} with ChPT at NLO using a resummed asymptotic formula.
To further check the values of $K_f^{FSE}$ the ETM Collaboration is generating further ensembles with the bare parameters equal to those of the ensemble cB211.25.48 (see Tab.~\ref{ensembles})
except for the volume. Using data from these ensembles we will be able to fix higher order
details and cross-check reliability of the chiral PT FSE correction formulae we employ.
 The result of the fit (\ref{f_pi_of_M_pi}) is plotted  in Fig.~\ref{fig:f_pi_of_M_pi}.
 Imposing  in the continuum limit at pion point that $M_\pi=134.80$~MeV \cite{Aoki:2016frl} and $f_\pi^{phen.}=130.41$~MeV \cite{PhysRevD.98.030001}
we find our preliminary estimate of the gradient flow scale $w_0$ and the other fit 
parameters (the coefficient describing lattice artefacts is compatible with zero)
\begin{equation}
w_0=0.1706(18) \,\mbox{fm}\,, \quad  f=122.31(18) \,\mbox{MeV}\,, \quad \bar\ell_4=4.3(1)\,, \quad P_4=-0.04(5).
\label{result_f_of_M_pi}
\end{equation}
The error quoted here and below in Eqs.~(\ref{res:f_K}), (\ref{res:f_D}) and (\ref{res:m_ud}) are only statistical.
Our value of the parameter $w_0$ is compatible with a previous estimate in the $N_f=2+1+1$ theory \cite{Bazavov:2015yea}. In future studies we plan to set the scale using the $\Omega$ baryon mass rather than $f_\pi^{phen}$.

 \vspace{-0.5cm}
\section{$f_K$ and $f_D$}\vspace{-0.3cm}

\begin{figure}[h]
   \vspace{-0.8cm} 
   \includegraphics[scale=0.5]{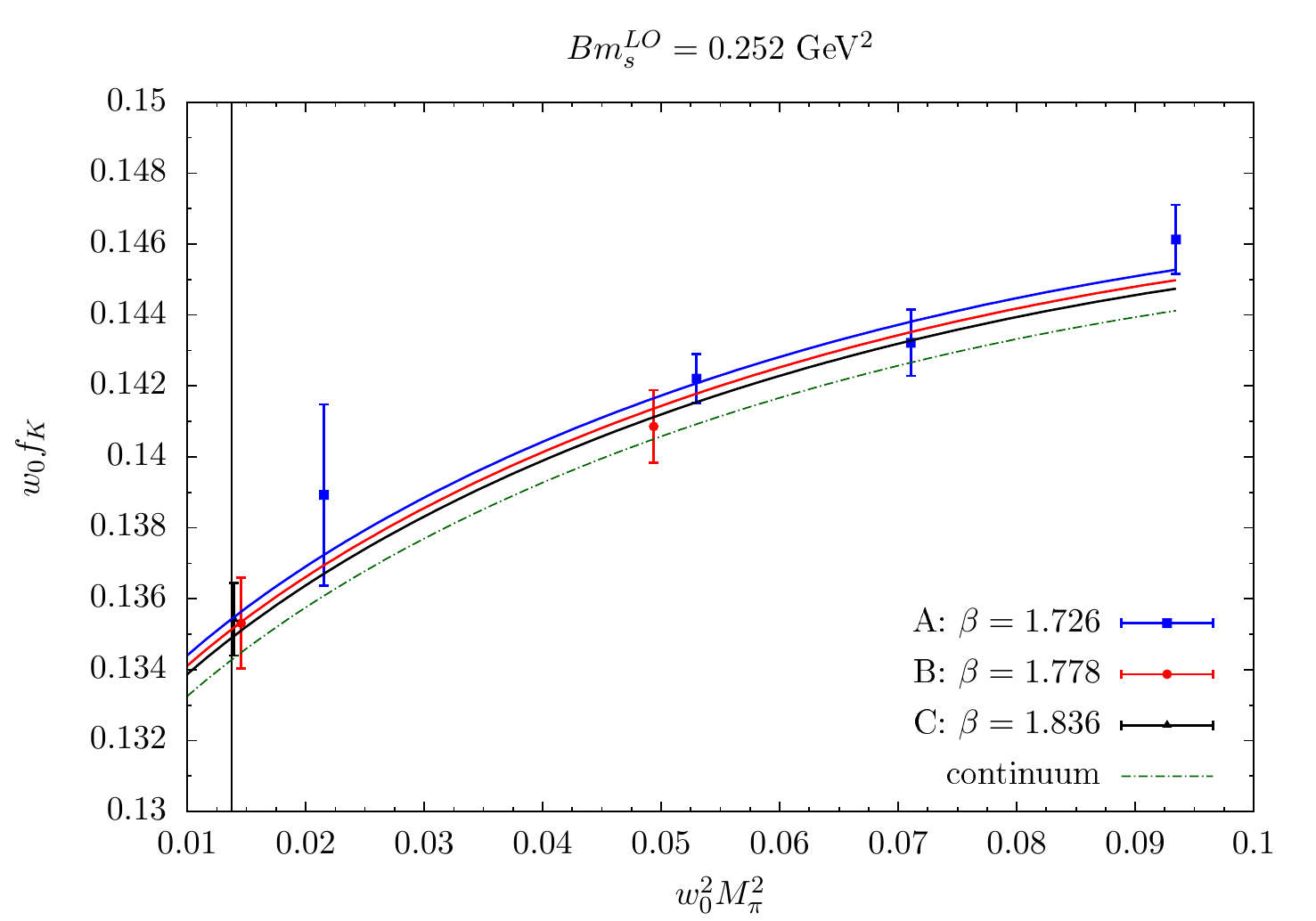}
  \includegraphics[scale=0.5]{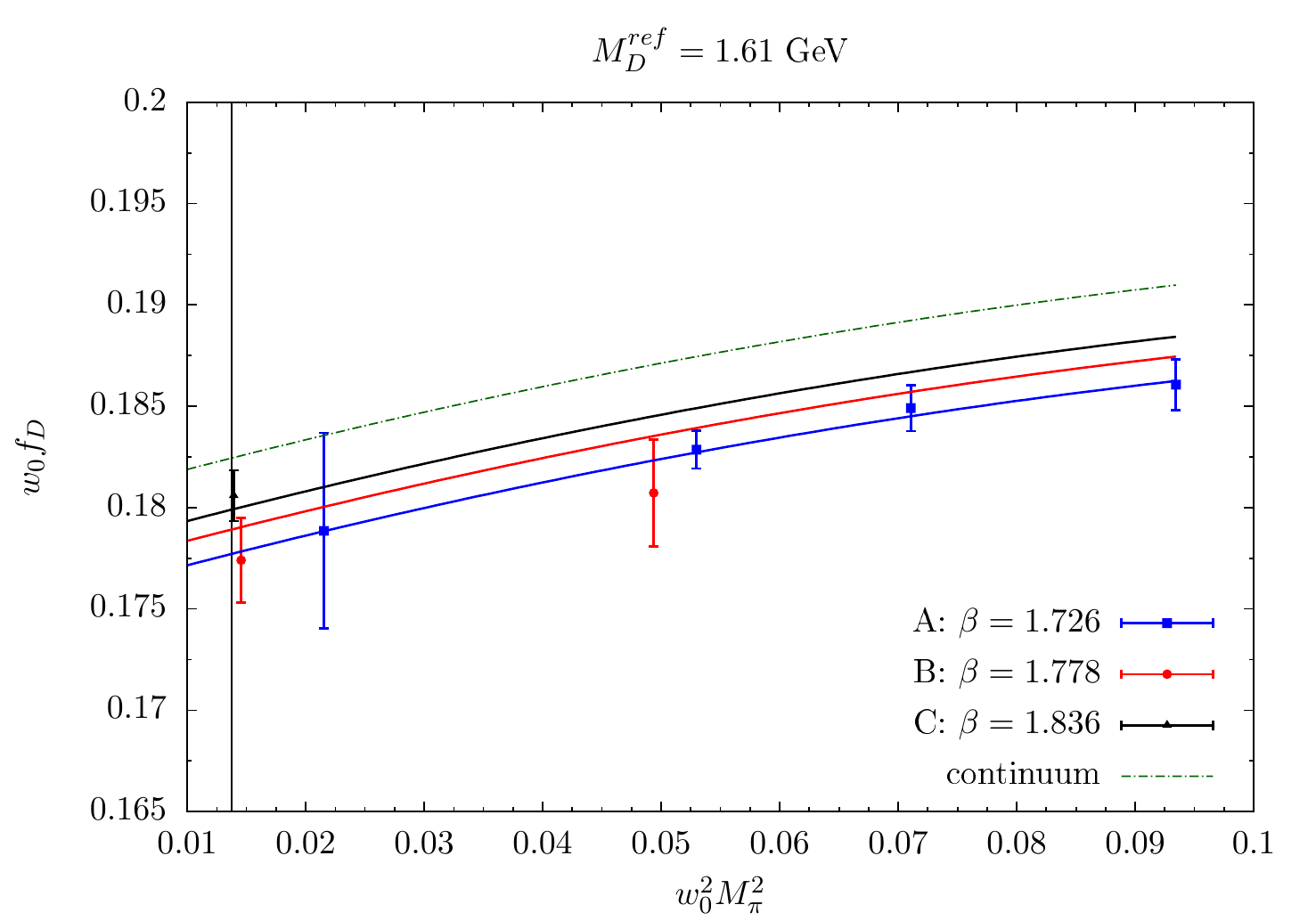}
\caption{Preliminary chiral and continuum fit of $f_K$ (left) and $f_D$ (right) as a function of $M_\pi$ in units of the gradient flow scale  $w_0$ \cite{Borsanyi2012}.
$f_K$ is plotted at a fixed $B m_s^{LO}=0.252$ GeV$^2$. 
$f_D$ is plotted at a fixed $M_D^{ref}=1.61$ GeV.
The fit functions we employ are given in Eq.~(\ref{f_K_of_M_pi})  for $f_K$ and
Eq.~(\ref{f_D_of_M}) for $f_D$.}
   \label{fig:f_K_D}
  \end{figure}

In order to extract the decay constant $f_K$ at the physical point we first perform a small
linear interpolation of our lattice data for each ensemble to three reference values of the quantity $B m_s^{LO}= M_K^2-M_\pi^2/2=0.186,\,0.212,\,0.252$ GeV$^2$ which 
at the leading order in ChPT is proportional to  the renormalised strange quark mass.
Then for each value of $B m_s^{LO}$ we extrapolate to the continuum limit and to the 
physical $M_\pi$ using our best fit to the data for $w_0 f_K$ according to the Ansatz
\begin{equation}\label{f_K_of_M_pi}
w_0 f_K\; = \; P_1^\prime\left[1-\frac{3}{4}\xi_\ell^M\log\xi_\ell^M+P_2^\prime\xi_\ell^M+ P_4^\prime a^2 \right]\,,
\end{equation}
where $P_1^\prime$, $P_2^\prime$ and $P_4^\prime$ depend on the specific value of $B{m}_s^{LO}$. 
In Fig.~\ref{fig:f_K_D} (left panel) we show the chiral and continuum extrapolation for the 
largest $B m_s^{LO}$.

Finally we interpolate linearly the tree values of $f_K$ obtained in the continuum and chiral fit to the physical  $ Bm_s^{LO}= M_K^2-M_\pi^2/2=0.233$ GeV$^2$, i.e. $M_K=494.2(4)$ MeV \cite{Aoki:2016frl}, obtaining preliminary estimates of $f_K$ (for the scale set as in Sect.~2) and the ratio $f_K/f_\pi$.
\begin{gather}\label{res:f_K}
 f_{K}=154(2) \; {\rm MeV} \,,\quad\quad
    f_{K}/f_{\pi}=1.182(16) \; {\rm MeV} \,,
\end{gather} 
in nice agreement with previous ETMC result \cite{Carrasco:2014poa}.\\
A similar analysis is performed to determine the decay constant of the $D$ meson. We first interpolate the data of each ensemble to certain reference masses $M_D^{ref}=1.61,\, 1.73,\, 1.84,\, 1.95 $ GeV (with scale from Sect.~2).
Then for each $M_D^{ref}$ mass value we extrapolate to the continuum limit and the physical $M_\pi$ using our best fit to the data according to the polynomial Ansatz
\begin{equation}\label{f_D_of_M}
w_0 f_{D}=P_1(1+P_2M_\pi^2+P_3M_\pi^4+P_4 a^2)\,.
\end{equation}
In Fig.~\ref{fig:f_K_D} (right panel) we show the chiral and continuum extrapolation for one 
typical reference $M_D^{ref}$ value.

Finally the four values of $f_D$ obtained for different $M_D^{ref}$ in the 
continuum and physical pion mass limits, are interpolated linearly to the physical
(isospin averaged)
$M_D^{exp}=1.867$ GeV \cite{PhysRevD.98.030001}, obtaining the preliminary  (with scale set as in Sect.~2)
\begin{equation}\label{res:f_D}
   f_{D}=215(6)\; {\rm MeV} \; .
\end{equation}  

 \vspace{-0.5cm}
\section{Renormalized light quark mass $m_{ud}$}
\begin{figure}[h]\centering
  \vspace{-0.1cm}
 \includegraphics[scale=0.5]{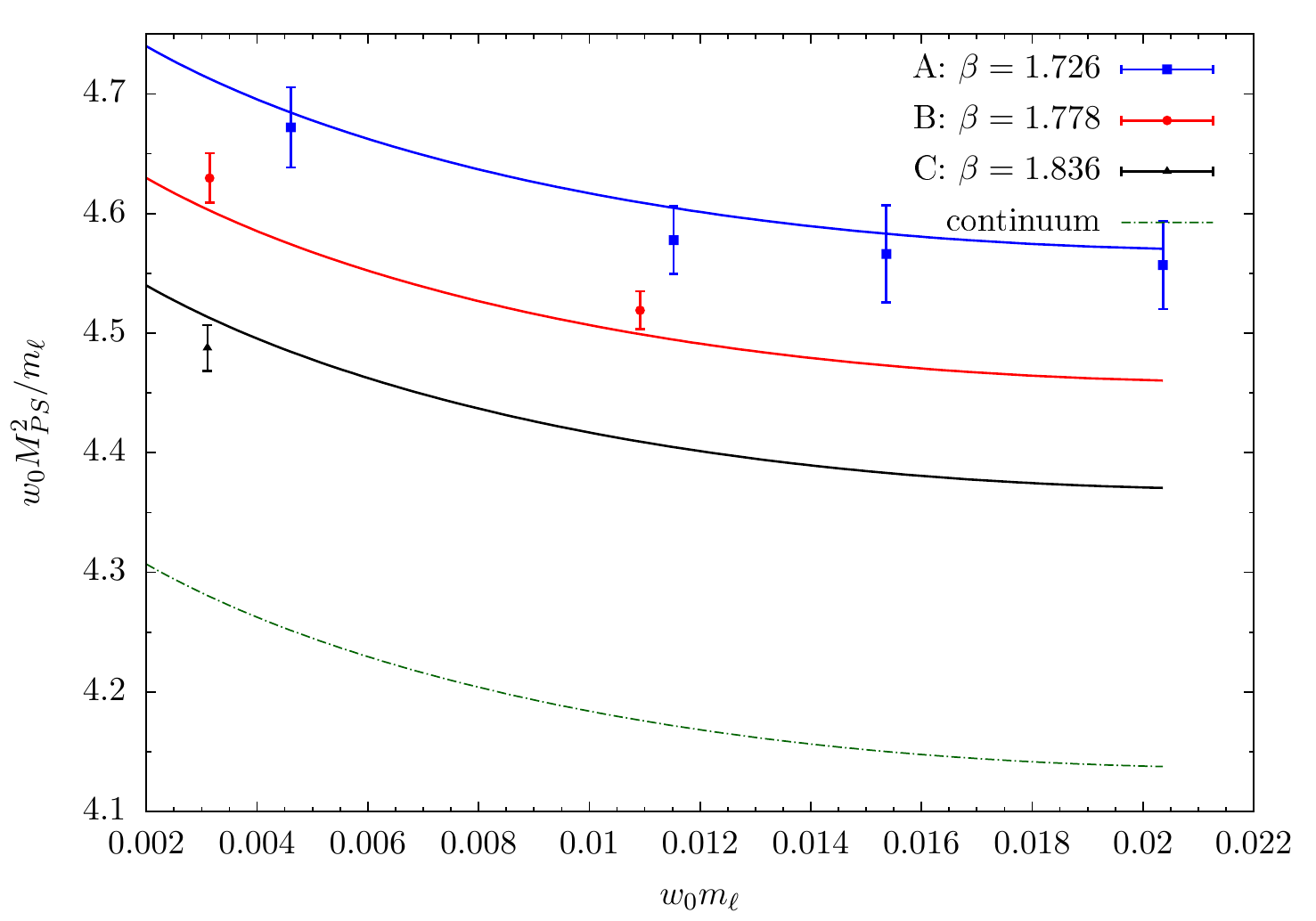}
  \includegraphics[scale=0.5]{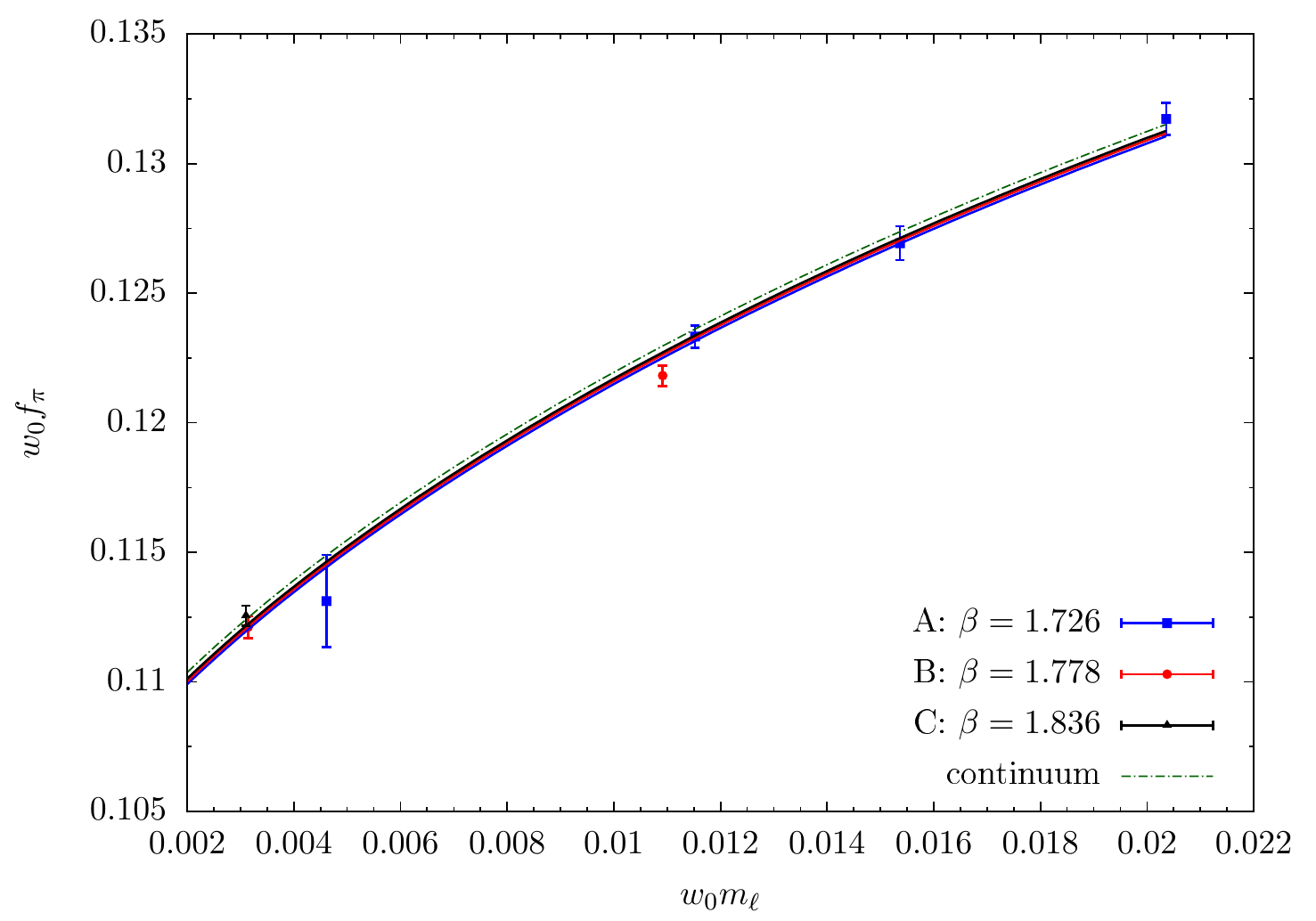}
  \vspace{-0.6cm}
 \caption{Preliminary chiral and continuum fits of $M_\pi$ (left) and $f_{\pi}$ (right) as a function of the renormalized quark mass $m_\ell$ in units of the gradient flow scale $w_0$ \cite{Borsanyi2012}.} 
   \vspace{-0.3cm}
 \label{fig:M_PS_f_PS_chiral}
 \end{figure}

\vspace{-0.3cm}
In this section we present our preliminary result for the average light quark renormalized 
mass $m_{ud}$. We fit the pion mass $M_\pi$ and the decay constant $f_\pi$ to the SU(2) ChPT 
formula 
\bea
     \label{eq:cptmpi2Ch}
\hspace{-0.4cm}(M_\pi  w_0 )^2  & = &\!\!\!\! 2(B w_0 )(m_\ell w_0 )\left[ 1 + \xi_\ell \log \xi_\ell  + P_1 \xi_\ell  +  P_2  \,a^2/w_0^2\right] K_{M^2}^{FSE} ~ ,  \\
    \label{eq:cptfpiCh}
\hspace{-0.4cm}    (f_\pi  w_0 ) & = &\!\!\!\! (f w_0 )\left[ 1 - 2\xi_\ell \log \xi_\ell  + P_3 \xi_\ell  +  P_4 \,a^2/w_0^2  \right] K_f^{FSE} ~ ,
 \eea
with the terms $\propto P_2$ and $\propto P_4$ describing the dominating lattice artifacts,
\begin{equation}
 \xi_\ell=2B_0m_\ell/(16 \pi^2 f^2)\,,\quad P_1=-\bar\ell_3-2\log{\left(M_\pi^{phys}/(4\pi f)\right)}\,,\quad
  P_3=2\bar\ell_4+4\log{\left(M_\pi^{phys}/(4\pi f)\right)}\nn
 \end{equation}
 and $m_\ell=\mu_\ell/Z_P$ the renormalized mass. The renormalization factors 
$Z_P$ have been calculated at each $g_0^2$ with $\sim 1\%$ percent accuracy in 
the RI'-MOM scheme and then converted (with N3LO accuracy) to the $\overline{MS}$ 
scheme using $N_f=4$ gauge ensembles generated for the purpose of evaluating
renormalization constants in mass-independent schemes.
The quantity $B,f,\bar{\ell_3},\bar{\ell_4},P_2,P_4$ are left as fit parameters 
and the  factor $K_{M^2/f}^{FSE}$ represents the correction for 
FSE as computed in Ref.~\cite{Colangelo:2005gd}.\\
In Fig.~\ref{fig:M_PS_f_PS_chiral} we show our chiral and continuum fits. 
Imposing as above $M_\pi=134.80$~MeV and  $f_\pi^{ phen}=130.41$~MeV we find 
the preliminary results
\begin{align}\label{res:m_ud}
& m_{ud}(\overline{MS},~2~{\rm GeV})=3.66(11) \; {\rm MeV} \,,
& \quad w_0=0.1703(18) \; {\rm fm} \,,
&\quad \bar\ell_3=2.9(2)\,,\\
& \quad \bar\ell_4=4.3(1)\, ,
&B=2539(78) \; {\rm MeV} \,,& \quad f=122.1(2) \; {\rm MeV} \, .
\end{align}
The values found for $w_0$, $\bar{\ell}_4$ and $f$ are compatible with those in Eq.~(\ref{result_f_of_M_pi}).
 
 \vspace{-0.3cm}
\section{Outlook and Acknowledgements}\vspace{-0.3cm}
We have discussed a first analysis of the gauge ensembles produced by ETMC with 
$N_f=2+1+1$ flavours of Wilson-clover twisted mass quarks. The results are preliminary
because a thorough study of the systematic uncertainties in  finite size corrections, in the
 chiral and continuum extrapolations and in the
computation of $m_{ud}$ renormalization constant is not included here. 
However already at this stage 
it is apparent that the physical pion mass point is safely reached and quite small lattice 
artifacts are found even for charmed observables.  

The computation of the correlators was carried out on the Marconi-KNL supercomputer 
at CINECA within the PRACE project~Pra17-4394. We thank all of ETMC for collaboration. This project was funded in part by the DFG as a project in the
Sino-German CRC110 (TRR110).
The authors gratefully acknowledge the Gauss Centre for Supercomputing
e.V. (www.gauss-centre.eu) for funding this project by providing
computing time on the GCS Supercomputers JUWELS~\cite{JUWELS} and
JURECA~\cite{jureca} at Jülich Supercomputing Centre (JSC).
F.S and S.S are supported by the Italian Ministry of Research (MIUR) under grant PRIN 20172LNEEZ. F.S is supported by INFN under GRANT73/CALAT.
 \vspace{-0.3cm}
\bibliographystyle{JHEP}
\bibliography{bibliography}

\end{document}